\documentclass[aps,prd,nofootinbib,showpacs,superscriptaddress, preprint]{revtex4-1}
\pdfoutput=1

\newcommand{\bmt}{\begin{pmatrix}}
\newcommand{\emt}{\end{pmatrix}}
\newcommand{\ba}{\begin{array}{c}}
\newcommand{\ea}{\end{array}}
\newcommand{\be}{\begin{equation}}
\newcommand{\ee}{\end{equation}}
\newcommand{\bea}{\begin{eqnarray}}
\newcommand{\eea}{\end{eqnarray}}

\newcommand{\bi}{\begin{itemize}}
\newcommand{\ei}{\end{itemize}}

\newcommand{\baz}{\begin{array}{cc}}
\newcommand{\besub}{\begin{subequations}}
\newcommand{\eesub}{\end{subequations}}

\newcommand{\mathsym}[1]{{}}

\newcommand{\bt}{\begin{tabular}}
\newcommand{\et}{\end{tabular}}

\newcommand{\benu}{\begin{enumerate}}
\newcommand{\eenu}{\end{enumerate}}
\newcommand{\ogw}{\Omega_\text{gw}}

\def\q2 {q^2}

\def\bt{\begin{table}}
\def\et{\end{table}}

\usepackage[utf8]{inputenc}
\usepackage{amsmath,amssymb}
\usepackage{epsfig}
\usepackage{comment}
\usepackage{graphicx}
\usepackage[usenames,dvipsnames]{color}
\usepackage{float}
\usepackage{bbold}
\usepackage{ wasysym }
\usepackage[colorlinks,citecolor=blue]{hyperref}
\usepackage{makecell}
\usepackage{color}
\usepackage{subfigure}

\begin{document}

\title{Imprint of PBH domination on gravitational waves generated by cosmic strings}



\author{Debasish Borah}
\email{dborah@iitg.ac.in}
\affiliation{Department of Physics, Indian Institute of Technology Guwahati, Assam 781039, India}

\author{Suruj Jyoti Das}
\email{suruj@iitg.ac.in}
\affiliation{Department of Physics, Indian Institute of Technology Guwahati, Assam 781039, India}

\author{Rishav Roshan}
\email{rishav.roshan@gmail.com}
\affiliation{Department of Physics, Kyungpook National University, Daegu 41566, Korea}
\affiliation{Center for Precision Neutrino Research, Chonnam National University, Gwangju 61186, Korea}

\author{Rome Samanta}
\email{samanta@fzu.cz}
\affiliation{CEICO, Institute of Physics of the Czech Academy of Sciences, Na Slovance 1999/2, 182 21 Prague 8, Czech Republic}

\begin{abstract}
 We study the effect of an ultra-light primordial black hole (PBH) dominated phase on the gravitational wave (GW) spectrum generated by a cosmic string (CS) network formed as a result of a high-scale $U(1)$ symmetry breaking. A PBH-dominated phase leads to tilts in the spectrum via entropy dilution and generates a new GW spectrum from PBH density fluctuations, detectable at ongoing and planned near-future GW detectors. The combined spectrum has a unique shape with a plateau, a sharp tilted peak over the plateau, and a characteristic fall-off, which can be distinguished from the one generated in the combination of CS and any other matter domination or new exotic physics. We discuss how ongoing and planned future experiments can probe such a unique spectrum for different values of $U(1)$ breaking scale and PBH parameters such as initial mass and energy fraction.
\end{abstract}

\maketitle

\section{Introduction}
While the standard model (SM) of particle physics has been the most successful phenomenological model explaining the elementary particles and their interactions except gravity, it can not explain several observed phenomena like neutrino mass, dark matter (DM), and baryon asymmetry. This has motivated the pursuit of exploring beyond standard model (BSM) scenarios for several decades. However, none of the particle physics-based experiments, including the ongoing Large Hadron Collider (LHC), have seen any signatures of BSM physics to date. This has garnered significant late interest in exploring alternative BSM-search strategies. Interestingly, the onset of the direct discovery of gravitational waves (GW) by LIGO-VIRGO collaboration \cite{LIGOScientific:2016aoc} has opened up an entirely new frontier to search for BSM physics with primordial gravitational waves. For example, BSM physics related to inflation, first-order phase transition, topological defects, and primordial black holes (PBH) may produce stochastic GW background within experimental sensitivity of ongoing and future planned experiments. This offers a synergic probe of BSM physics and could be the only realistic probe in some scenarios that are out of reach of particle physics experiments in the foreseeable future. See a recent review \cite{Caldwell:2022qsj} discussing GW signatures of BSM physics.

Many well-motivated BSM scenarios lead to the formation of topological defects like cosmic strings and domain walls \cite{Vilenkin:1984ib}. In particular, Abelian gauge extensions of the SM, with several particle physics motivations \cite{Langacker:2008yv}, lead to the formation of cosmic strings (CS) after spontaneous symmetry breaking \cite{Kibble:1976sj, Nielsen:1973cs}. These CS can generate stochastic GW background with a characteristic spectrum which can be observed at near future GW detectors if the symmetry breaking scale is sufficiently high \cite{Vilenkin:1981bx,Turok:1984cn}, far outside the reach of the direct probe at experiments like the LHC. After the recent findings by NANOGrav collaboration suggesting a stochastic common spectrum process across many pulsars \cite{NANOGrav:2020bcs}, CS as a source of primordial GWs gained a great deal of attention \cite{Goncharov:2021oub,Ellis:2020ena, Blasi:2020mfx,Samanta:2020cdk, Hindmarsh:2022awe}. While a scale-invariant GW spectrum at higher frequencies is a typical signature of CS, any deviation from this spectrum can signify additional new physics or non-standard cosmological epochs in the early universe. For example, early matter domination (EMD) can lead to spectral breaks in the CS-generated GW spectrum \cite{Cui:2017ufi,Cui:2018rwi,Gouttenoire:2019kij}. Such an EMD phase can arise due to long-lived particle \cite{Allahverdi:2021grt, Borah:2022byb}, PBH, or a combination of both \cite{Barman:2022gjo, Borah:2022vsu}. Similarly, new physics, which might incorporate an additional source of GW, can also lead to distortions in the CS-generated GW spectrum. For instance, a first-order phase transition (FOPT) induced GWs together with the CS generated GW spectrum \cite{ZhouRuiYu:2020bbs, Ferrer:2023uwz} can have distinct features which can be detectable at near future experiments if the FOPT is a supercooled one \cite{Ferrer:2023uwz}. While the effect of EMD due to PBHs on CS-generated GW spectrum has been studied in several works \cite{Datta:2020bht, Samanta:2021mdm, Borah:2022iym, Borah:2022vsu, Ghoshal:2023sfa}, not much attention has been paid to the additional effect caused by GWs originating from PBH themselves except for \cite{Borah:2022iym}, where the individual GW contributions from CS and PBH were studied in the context of superheavy dark matter and high scale leptogenesis. 

Motivated by these, we perform a general study on  the effect of an ultralight PBH-dominated epoch on the CS-generated GWs  and present the result for the combined spectrum that also includes, unlike any other EMDs, features of GWs caused by PBH themselves.  Among the various ways PBHs associate with GWs, we consider the one induced due to PBH density fluctuations \cite{Carr:2020gox,Inomata:2019ivs,Papanikolaou:2020qtd, Domenech:2021wkk, Domenech:2020ssp, Domenech:2021ztg}. For ultra-light PBH domination, this GW contribution remains detectable in ongoing and near-future experiments. We show that  the combined  spectrum  exhibits a plateau, a red tilt followed or  preceded by a sharp blue tilt. Besides being distinct from any other spectrum, e.g., a CS-generated GW spectrum with an EMD caused by a fermion/scalar field \cite{Borah:2022byb,ls1,ls2}, the overall spectrum offers characteristic features that can be detected either with a single detector or  with multiple detectors in a complementary way.

This paper is organised as follows. In section \ref{sec1} and section \ref{sec2}, we briefly summarise the GW production from cosmic strings and PBH density fluctuations, respectively. In section \ref{sec3}, we discuss the combined GW spectrum and finally conclude in section \ref{sec4}.

\section{Gravitational Waves from Cosmic Strings}
\label{sec1}
If the vacuum manifold of a symmetry group after spontaneous symmetry breaking is not simply connected, one-dimensional topological defects namely, cosmic strings are formed \cite{Kibble:1976sj, Nielsen:1973cs}. Such a possibility naturally arises in popular $U(1)$ extensions of the SM. If this $U(1)$ symmetry is a gauged one, then the cosmic string loops lose their energy dominantly in the form of GW radiation, as suggested by numerical simulations based on Nambu-Goto action \cite{Ringeval:2005kr,Blanco-Pillado:2011egf}.  For a sufficiently high $U(1)$ symmetry breaking scale ($\Lambda_{CS}\gtrsim 10^9$ GeV), the resulting GW background is detectable at ongoing and near future GW experiments. This makes CS generated GW an outstanding probe of super-high scale BSM physics \cite{Buchmuller:2013lra, Dror:2019syi, Buchmuller:2019gfy,King:2020hyd, Fornal:2020esl,  Buchmuller:2021mbb,Guedes:2018afo,Sousa:2020sxs, Masoud:2021prr,Lazarides:2021uxv, Afzal:2022vjx, Borah:2022byb,Lazarides:2022jgr,Lazarides:2022spe,Maji:2022jzu} which can not be probed directly at terrestrial experiments. The key parameter related to CS is their normalised tension $G \mu\sim G \Lambda_{\rm CS}^2$ with $G$ being Newton's constant. In the absence of thermal friction leading to dampening of the motion of a long-string network \cite{Vilenkin:1991zk}, shortly after formation, the network oscillates (at $t_{\rm osc}$) and enters the scaling regime \cite{Blanco-Pillado:2011egf,Bennett:1987vf,Bennett:1989ak} which is an attractor solution of two competing dynamics namely, the stretching of the long-string correlation length due to cosmic expansion and the fragmentation of the long strings into loops which oscillate to produce particle radiation or GW \cite{Vilenkin:1981bx,Turok:1984cn,Vachaspati:1984gt}. 

A set of normal-mode oscillations with frequencies $f_k=2k/l$ constitute the total energy loss from a cosmic string loop, where the mode numbers are denoted by $k=1,2,3....\infty$. The GW energy density parameter can, therefore, be defined as $\Omega_{\rm GW}(t_0,f)=\sum_k\Omega_{\rm GW}^{(k)}(t_0,f)$, with $t_0$ being the present time and $f\equiv f(t_0)= f_k a(t_0)/a(t)$. The corresponding GW energy density at the present epoch for the mode $k$ can be computed with the integral \cite{Blanco-Pillado:2013qja} 
\begin{align}
 \Omega_{\rm GW}^{(k)}(t_0,f)=\frac{2k G\mu^2\Gamma_k}{f \rho_c}\int_{t_{osc}}^{t_0}dt\left[\frac{a(t)}{a(t_0)}\right]^5 n\left(t,l_k\right),\label{gwint}
\end{align}
where $n\left(t,l_k\right)$ is a scaling loop number density which can be computed analytically using Velocity-dependent-One-Scale (VOS) \cite{Martins:1996jp,Martins:2000cs,Sousa:2013aaa,Auclair:2019wcv} model, $\rho_c$ is the critical energy density of the universe and $\Gamma_k=\frac{\Gamma k^{-\delta}}{\zeta(\delta)}$ depends on the small-scale structures in the cosmic string loops such as cusps ($\delta=4/3$) and kinks ($\delta=5/3$). Here, we consider only cusps to compute the GW spectrum; a similar analysis can also be done considering kinks. While it is clear that for higher modes, contributions to the GWs from cusps are dominant, the number of cusps and kinks per loop oscillation cannot be straightforwardly determined with numerical simulations, which do not include gravitational wave back-reaction in general. A preference for cusps over kinks comes from the so-called smoothing mechanism, where the kinky loops are expected to be smoothed out due to the gravitational wave back-reaction \cite{Blanco-Pillado:2015ana}. This mechanism, however, has been challenged in \cite{Wachter:2016hgi,Wachter:2016rwc}. Therefore, we would like to place the consideration of cusps over kinks in this article as a choice rather than a preference. 
 The integration in Eq.\ref{gwint} may be subjected to another constraint: the critical size of the loops below which particle production is dominant. This constraint corresponds to a lower bound on the time GW radiation becomes effective \cite{Matsunami:2019fss,Auclair:2019jip}. We carefully consider this constraint in the numerical analysis.

A characteristic feature of the GW spectrum generated by CS is the flat plateau due to loop formation and decay during radiation dominated phase of the universe, with an amplitude given by 
\begin{align}
\Omega_{\rm GW}^{(k=1),{\rm plateau}}(f)=\frac{128\pi G\mu}{9\zeta(\delta)}\frac{A_r}{\epsilon_r}\Omega_r\left[(1+\epsilon_r)^{3/2}-1\right], \label{flp1}
\end{align}
where $\epsilon_r=\alpha/\Gamma G\mu$ with $\alpha$ the initial (at $t=t_i$) loop size parameter, $\Omega_r\simeq 9\times 10^{-5}$ and $A_r=5.4$ \cite{Auclair:2019wcv}. In our work, we have considered $\alpha \simeq 0.1$\,\cite{Blanco-Pillado:2013qja,Blanco-Pillado:2017oxo} (as indicated by numerical simulations),  $\Gamma\simeq 50$\,\cite{Vachaspati:1984gt,Blanco-Pillado:2013qja,Blanco-Pillado:2017oxo}, and cosmic microwave background (CMB) constraint $G\mu\lesssim 10^{-7}$ \cite{Charnock:2016nzm} which lead to $\alpha \gg \Gamma G \mu$. In this limit, Eq.\eqref{flp1} implies $\Omega_{\rm GW}^{(k=1)}(f)\sim \Lambda_{\rm CS}$, a property that makes models with larger symmetry breaking scales more testable with GWs from CS. While this proportionality is robust for large loops ($\alpha\simeq 0.1$), the overall magnitude of $\Omega_{\rm GW}^{(k=1)}(f)$ in Eq.(\ref{flp1}) obtained from the VOS model differs from that of the numerical simulations by order of magnitude. This is related to the fact that the VOS model considers all the loops contributing to the GWs to be created with the same initial size. On the other hand, numerical simulations find only 10$\%$ of energy from the long strings going to the large loops ($\alpha\simeq 0.1$), which contribute to the GWs; the rest is transferred to the smaller loops in the form of kinetic energy which redshifts away and does not contribute to the GWs  \cite{Blanco-Pillado:2013qja}. Therefore, a factor $\mathcal{F}_\alpha\simeq 0.1$ is introduced in Eq.(\ref{flp1}) (which is equivalent to normalizing loop number density in the VOS model, where the normalization factor is $\mathcal{F}_\alpha$, see, e.g., sec.V of \cite{Sousa:2020sxs} for a comprehensive description of normalization) such that the string network evolution and its properties can be described consistently with VOS model, and at the same time, with the numerical simulations.


We shall see later in this article that in the presence of a matter-dominated epoch before the most recent radiation domination, the plateau breaks at a frequency $f_\Delta$ and the spectrum falls as $\Omega_{\rm GW} (f>f_\Delta) \sim f^{-1}$ for the fundamental mode of loop-oscillations. The frequency $f_\Delta$ carries the information regarding the end of the matter domination and, potentially, other properties of the object/field associated with the matter domination.
\section{Gravitational Waves from primordial black holes}
\label{sec2}
Primordial black holes (A recent review can be found in \cite{Carr:2020gox}), originally proposed by Hawking \cite{Hawking:1974rv, Hawking:1975vcx} can have many interesting cosmological consequences \cite{Chapline:1975ojl, Carr:1976zz} including detection prospects via GW. While PBH can have a wide range of masses, we are interested in the ultra-light regime in which they evaporate by emitting Hawking radiation \cite{Hawking:1974rv, Hawking:1975vcx} before the BBN epoch and hence are less constrained from cosmological and astrophysical observations. In the early universe, PBH can be formed in a variety of ways like, from inflationary perturbations \cite{Hawking:1971ei, Carr:1974nx, Wang:2019kaf, Byrnes:2021jka, Braglia:2022phb}, first-order phase transition \cite{Crawford:1982yz, Hawking:1982ga, Moss:1994iq, Kodama:1982sf, Baker:2021nyl, Kawana:2021tde, Huang:2022him, Hashino:2021qoq, Liu:2021svg}, the collapse of topological defects \cite{Hawking:1987bn, Deng:2016vzb} etc. Here we remain agnostic about such a production mechanism and assume an initial abundance of PBH (of Schwarzschild type) with a monochromatic mass function. PBH is assumed to be formed in the era of radiation domination after inflation, and its abundance is characterized by a dimensionless parameter $\beta$ defined as 
\bea
\beta=\frac{\rho_\text{BH}(T_\text{in})}{\rho_\text{R}(T_\text{in})},
\eea
where $\rho_\text{BH}(T_\text{in})$ and $\rho_\text{R}(T_\text{in})$ represent the initial PBH energy density and radiation energy density respectively while $T_\text{in}$ denotes the temperature at the time of PBH formation. Such PBH can lead to an early matter domination phase if its initial energy fraction $\beta$ exceeds a critical value defined by \cite{Masina:2020xhk}, 
\begin{equation}
\beta<\beta_\text{c}\equiv \gamma^{-1/2}\,\sqrt{\frac{\mathcal{G}\,g_{\star,H}(T_\text{BH})}{10640\,\pi}}\,\frac{M_\text{pl}}{m_\text{in}}\,,
\label{eq:pbh-ev-rad}
\end{equation}
where
\begin{equation}
{g_{\star,H}}(T_\text{BH})\equiv\sum_i\omega_i\,g_{i,H} e^{-\frac{M_{BH}}{\beta_i M_i}}\,; g_{i,H}=
    \begin{cases}
        1.82
        &\text{for }s=0\,,\\
        1.0
        &\text{for }s=1/2\,,\\
        0.41
        &\text{for }s=1\,,\\
        0.05
        &\text{for }s=2\,,\\
    \end{cases}
    \label{eqn:gsh}
\end{equation}
with $\omega_i=2\,s_i+1$ for massive particles of spin $s_i$, $\omega_i=2$ for massless species with $s_i>0$ and $\omega_i=1$ for $s_i=0$. $\beta_s$ = 2.66, 4.53, 6.04, 9.56  for $s_i$ = 0,1/2,1,2 respectively, whereas $M_{i}= \frac{1}{8\pi G m_{i}}$, where $m_{i}$ is the mass of the $i^\text{th}$ species \cite{Masina:2020xhk}. $\gamma\simeq0.2$ denotes a numerical factor related to the uncertainty of the PBH formation, $\mathcal{G}\sim 4$ is the grey-body factor with $m_\text{in}$ and $T_\text{BH}$ denoting the mass of PBH at the time of formation and instantaneous Hawking temperature of PBH respectively.

As per GW signatures of PBH are concerned, there are variety of ways this can happen. The evaporation of PBH itself can produce gravitons which might constitute an  ultra-high frequency  GW spectrum \cite{Anantua:2008am}. PBH can also form mergers, leading to GW emission \cite{Zagorac:2019ekv}. In addition, the scalar perturbations leading to the formation of PBH can induce GWs at second-order \cite{Saito:2008jc}, which can be further enhanced during the PBH evaporation \cite{Inomata:2020lmk}. Finally, the inhomogeneity in the distribution of PBH may also induce GW at second order, as recently studied in  Ref. \cite{Papanikolaou:2020qtd, Domenech:2020ssp, Domenech:2021wkk}. We concentrate on this last possibility particularly because it can be tackled while being agnostic about PBH formation history, and the resulting GW spectrum can be probed at ongoing as well as planned near-future GW detectors.

The distribution of PBHs after they are formed is random and follows Poissonian statistics \cite{Papanikolaou:2020qtd}. When PBHs dominate the universe's energy density, these inhomogeneities induce GWs at the second order, which are enhanced further during PBH evaporation \cite{Domenech:2020ssp}. The dominant contribution to the GW amplitude observed today can be written  as \cite{Domenech:2020ssp, Borah:2022iym, Barman:2022pdo}
\begin{equation}
    \ogw(t_0,f)\simeq \ogw^{\rm peak}\left(\frac{f}{f_{\rm peak}}\right)^{11/3}\Theta
\left(f_{\rm peak}-f\right),\label{eqn:omgwpbh}
\end{equation}
where
\begin{equation}
    \ogw^{\rm peak}\simeq 2\times 10^{-6} \left(\frac{\beta}{10^{-8}}\right)^{16/3}\left(\frac{m_{\text{in}}}{10^7 \rm g}\right)^{34/9}.\label{eqn:omgpeakpbh}
\end{equation}
The GW spectrum has an ultraviolet cutoff at frequencies corresponding to comoving scales representing the mean separation between PBH \cite{Papanikolaou:2020qtd, Domenech:2020ssp}, which is given by 
\begin{equation}
    f_{\rm peak}\simeq 1.7\times 10^3\,{\rm Hz}\,\left(\frac{m_{\text{in}}}{10^4 \rm g}\right)^{-5/6}.\label{eqn:fpkpbh}
\end{equation}
Although we shall use the Eq.(\ref{eqn:omgwpbh})-(\ref{eqn:fpkpbh}) for numerical computation, we note that the amplitude of the induced GWs is sensitive to the width of the PBH mass function \cite{Papanikolaou:2022chm,Inomata:2020lmk,Kozaczuk:2021wcl}. In addition, density perturbations can become non-linear during the PBH-dominated era, leading to a suppression of the spectrum \cite{Kozaczuk:2021wcl}.

\section{the Combined GW Spectrum, detection prospects, and realistic BSM physics scenarios}
\label{sec3}

{\it The combined spectrum:} As mentioned earlier, ultra-light PBHs with $\beta > \beta_c$ can affect the CS-generated GW spectrum in two ways: introducing spectral break due to PBH domination plus evaporation and  introducing a new GW source from density fluctuations. Due to the PBH-induced early matter domination before BBN, the plateau region of the CS-generated GW spectrum gets broken \cite{Cui:2017ufi,Cui:2018rwi,Gouttenoire:2019kij} at a turning point frequency (TPF) $f_\Delta$ which can be determined by the time say $t_\Delta$ (or equivalently the PBH evaporation temperature $T_\Delta$) at which the early matter domination ends and the standard radiation era begins. The frequency $f_\Delta$ (present-day) is equivalent to twice the inverse length $l_M^{-1}$ of a loop  (with initial size $\alpha t_\Delta$)  corresponding to the maximal emission of GWs at $t_M >t_\Delta$. A higher value of $f_\Delta$ would imply the loop was produced earlier--radiation domination began earlier. A loop created at a time $t_\Delta$ contributes maximally to Eq. \eqref{gwint} when it reaches its half-life, i.e., $l_{M}(t_M)=\frac{  \alpha t_\Delta}{2}$ \cite{Cui:2018rwi}. Therefore, the frequency observed today for GWs emitted at $t_M$ is given by
\bea
f_\Delta=\frac{4}{\alpha t_\Delta} \frac{a_M}{a_0}=\sqrt{\frac{8}{\alpha \Gamma G \mu}} t_\Delta^{-1/2}t_{\rm eq}^{1/6} t_0^{-2/3}\simeq \sqrt{\frac{8 z_{\rm eq}}{\alpha\Gamma G\mu}}\left(\frac{t_{\rm eq}}{t_{\Delta}}\right)^{1/2} t^{-1}_0,\label{tpf1}
\eea
where $a_M$ is the scale factor at the time $t_M=\frac{\alpha t_\Delta}{2\Gamma G\mu}$, $t (z)_{\rm eq}$ is the time (redshift $\sim 3400$) corresponding to the standard matter-radiation equality, and $t_0$ is the present time. Eq.(\ref{tpf1}) can be conveniently recast in terms of PBH evaporation and present-day temperatures as
\begin{align}
 f_\Delta&=\sqrt{\frac{8}{z_{\rm eq}\alpha\Gamma G\mu}}\left(\frac{g_*(T_\Delta)}{g_*(T_0)}\right)^{1/4}\frac{T_\Delta}{T_0}t_0^{-1} \nonumber \\ &\simeq 0.656 \left(\frac{\alpha}{0.1}\right)^{-1/2}\left(\frac{\Gamma}{50}\right)^{-1/2}\left(\frac{G\mu}{10^{-13}}\right)^{-1/2}\left(\frac{g_*(T_\Delta)}{g_*(T_0)}\right)^{1/4} \left(\frac{T_\Delta}{\text{GeV}} \right)~{\rm Hz},\label{eq:TPF}
\end{align}
which implies, e.g., if the PBHs evaporate at  $T_\Delta= 1$ TeV and $1$ MeV, the plateau of the GW spectrum  gets broken at $f_\Delta \simeq 1558$ Hz and $0.9$ mHz, respectively for typical values of $G\mu$, $\Gamma$, $\alpha$ in Eq.(\ref{eq:TPF}). Beyond $f_\Delta$, the spectrum goes as $\Omega_{\rm GW}\sim f^{-1}$ for $k=1$ mode (when infinite modes are summed, $\Omega_{\rm GW}\sim f^{-1/3}$\cite{Blasi:2020wpy,Gouttenoire:2019kij,Datta:2020bht, Gouttenoire:2019rtn}). This fall persists in general because, for considerably long PBH domination, contributions from the loops produced in the PBH-dominated era and before, are negligible or show up at higher frequencies with subdominant amplitude which we do not discuss in detail in this paper. Thus, the primary first effect or outcome due to the PBH-domination is a flat GW spectrum turning red beyond $f_\Delta$.

When the second effect, i.e., the contribution of the GWs from PBH density fluctuation, is considered, the combined spectrum shows a blue tilted peak with a sharp cut-off at $f_{\rm peak}$, in addition to the plateau and the red tilt feature discussed above. Depending on the value of  $f_{\rm peak}$ and $f_{\Delta}$, we consider three distinct cases to discuss our results: (i) $f_{\rm peak} > f_{\Delta}$, (ii) $f_{\rm peak} < f_{\Delta}$ and (iii) $f_{\rm peak} \sim f_{\Delta}$.
\begin{figure}
    \centering
  \includegraphics[scale=.42]{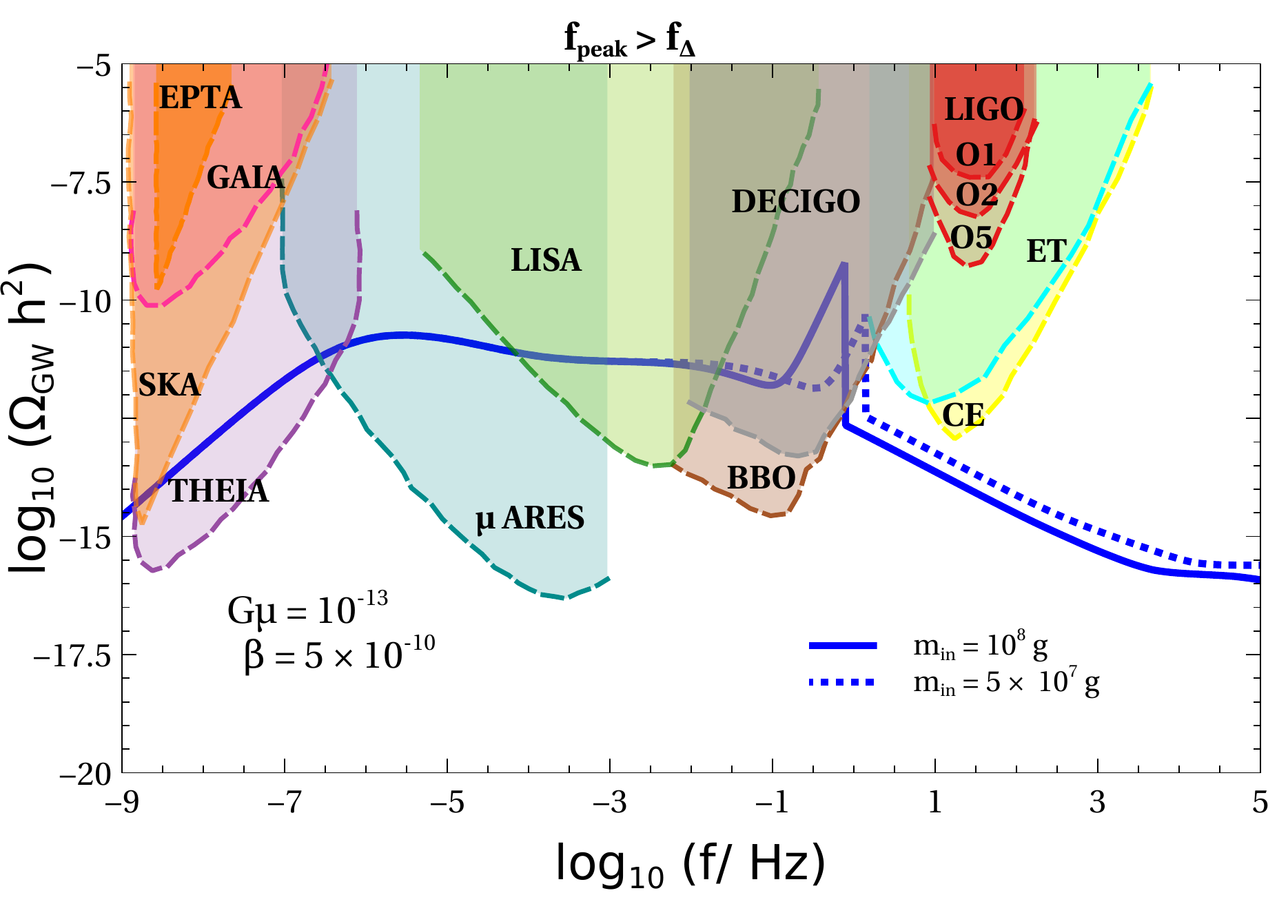}~~ \includegraphics[scale=.42]{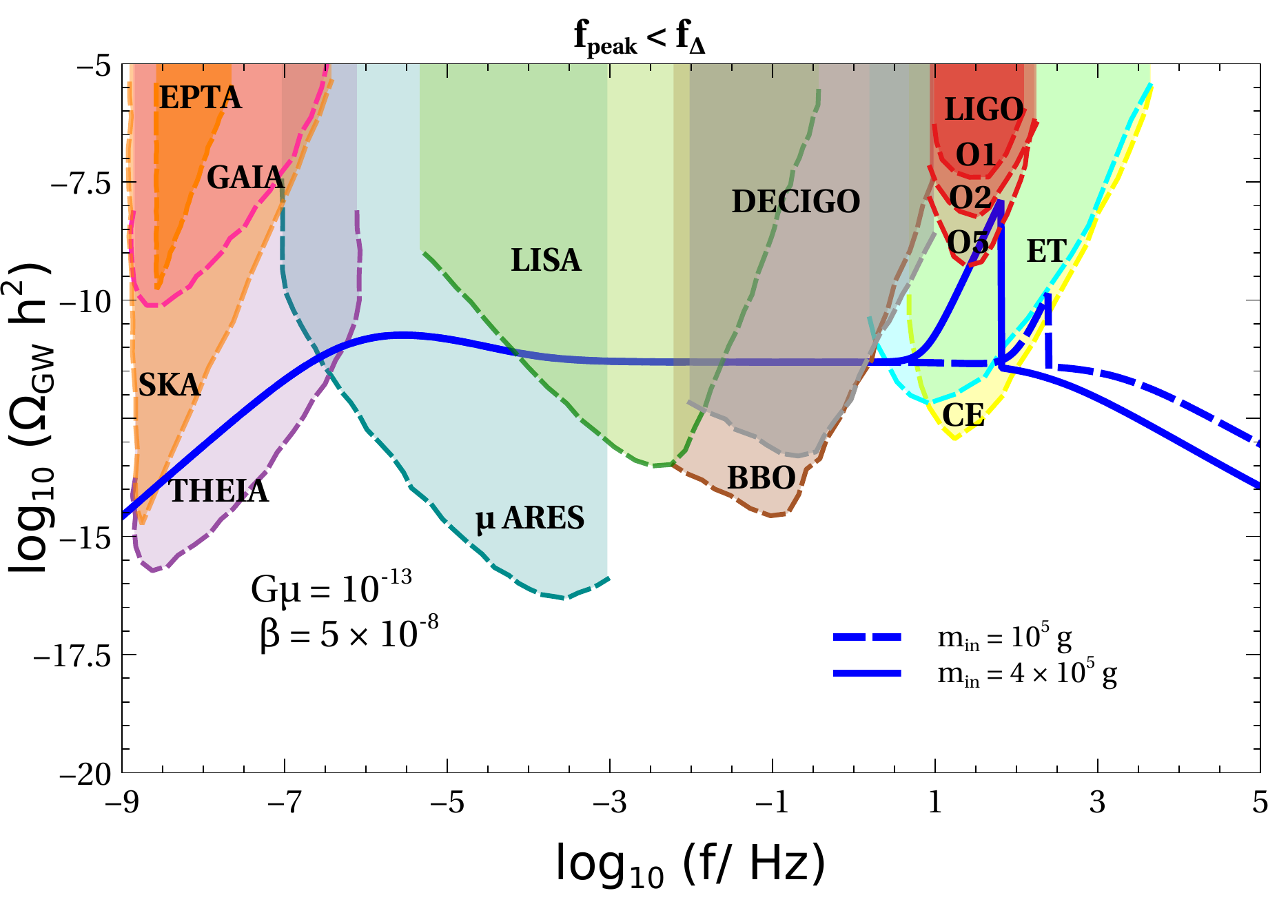}
    \caption{Combined GW spectra from cosmic strings and PBH density fluctuations, for a fixed value of $G\mu$ and for different values of initial PBH mass $m_{\rm in}$. In the left panel, we have $f_{\rm peak} > f_{\rm \Delta}$, whereas in the right panel, $f_{\rm peak} < f_{\rm \Delta}$. }
    \label{fig1}
\end{figure}

\begin{figure}
    \centering
  \includegraphics[scale=.42]{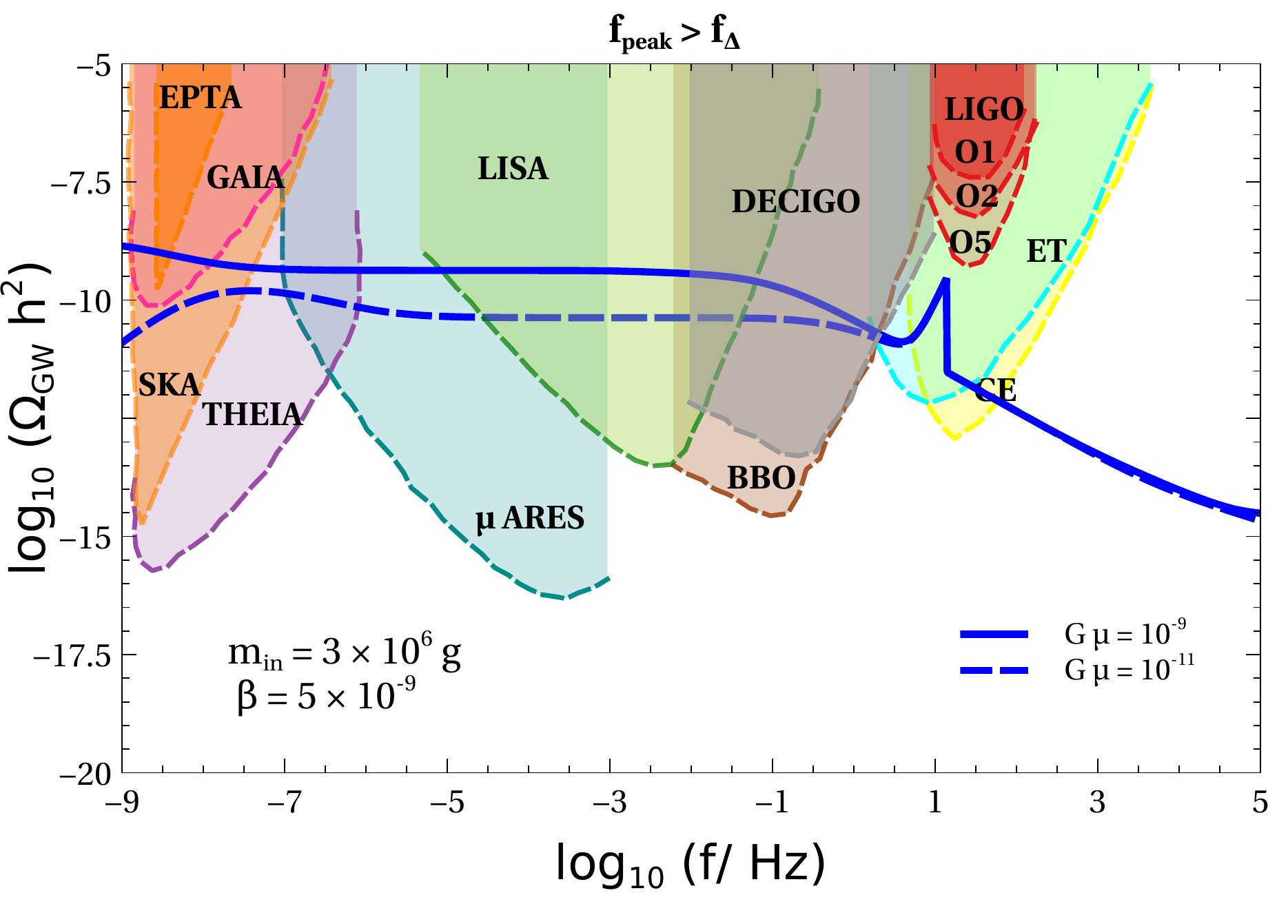}~~ \includegraphics[scale=.42]{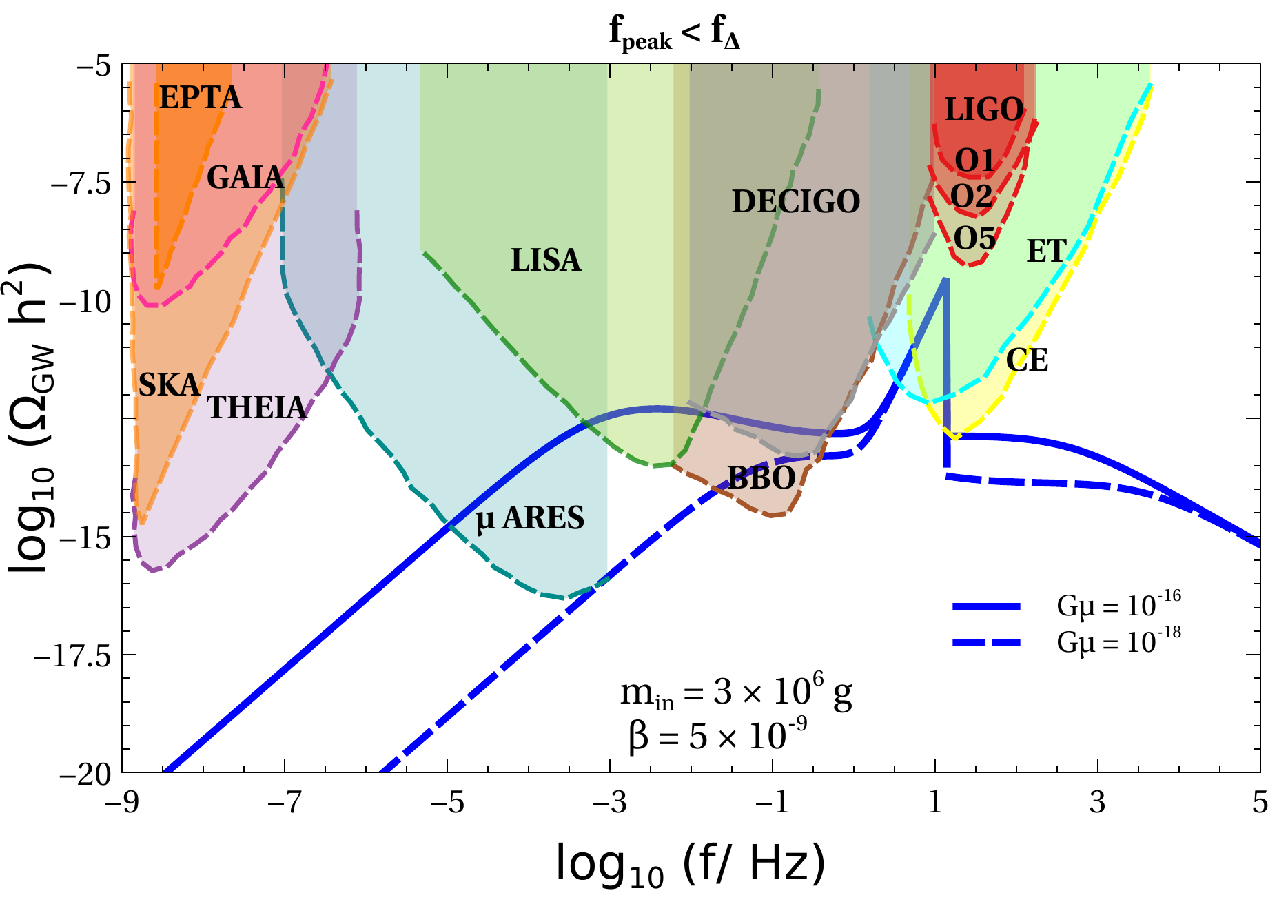}
    \caption{Combined GW spectra from cosmic strings and PBH density fluctuations, for a fixed value of initial PBH mass $m_{\rm in}$, and for different values of $G\mu$. In the left panel, we have $f_{\rm peak} > f_{\rm \Delta}$, whereas in the right panel, $f_{\rm peak} < f_{\rm \Delta}$.}
    \label{fig2}
\end{figure}

\begin{figure}
    \centering
  \includegraphics[scale=.55]{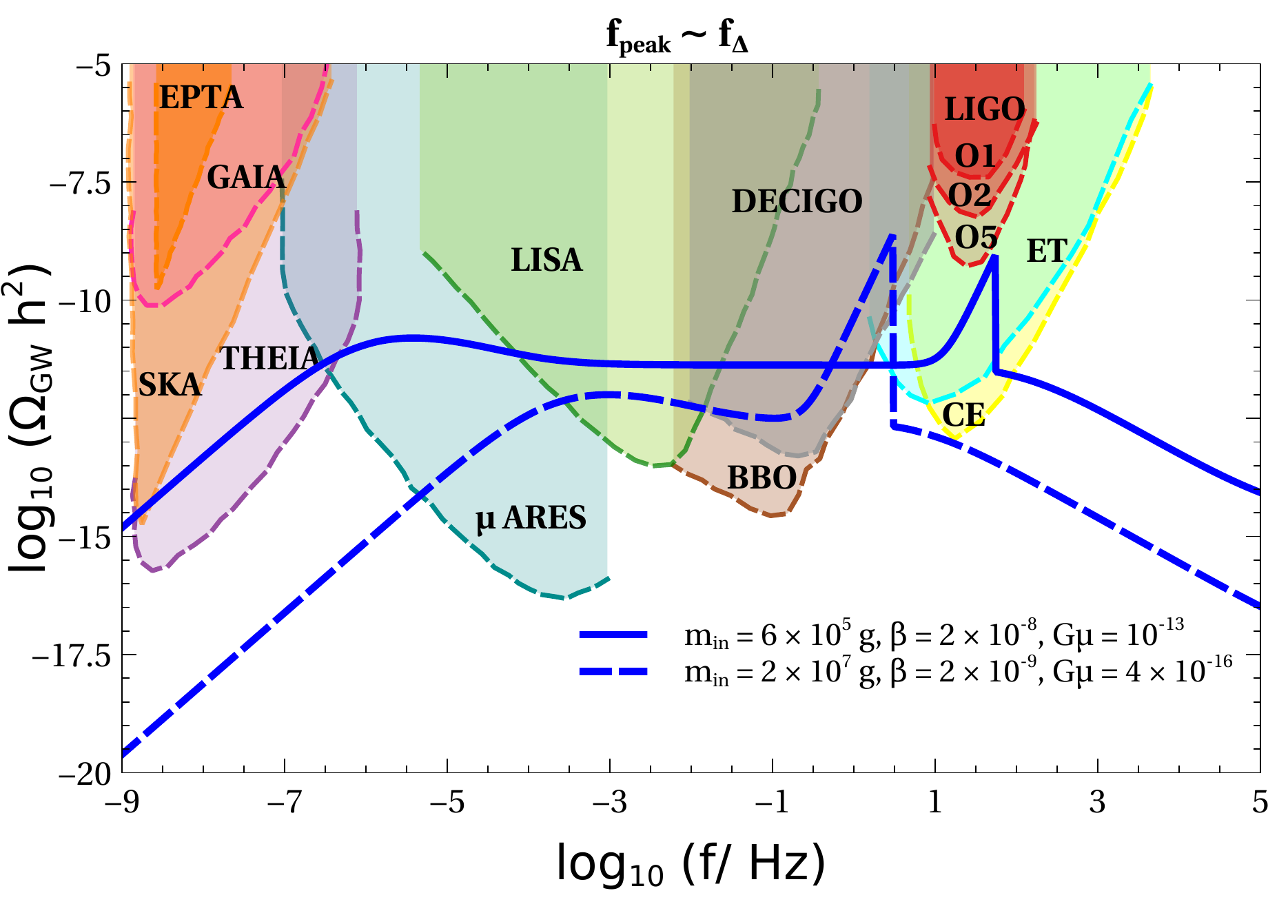}~~ 
    \caption{Combined GW spectra from cosmic strings and PBH density fluctuations, choosing benchmark values for which $f_{\rm peak} \sim f_{\rm \Delta}$.}
    \label{fig3}
\end{figure}
\begin{figure}
    \centering
  \includegraphics[scale=.42]{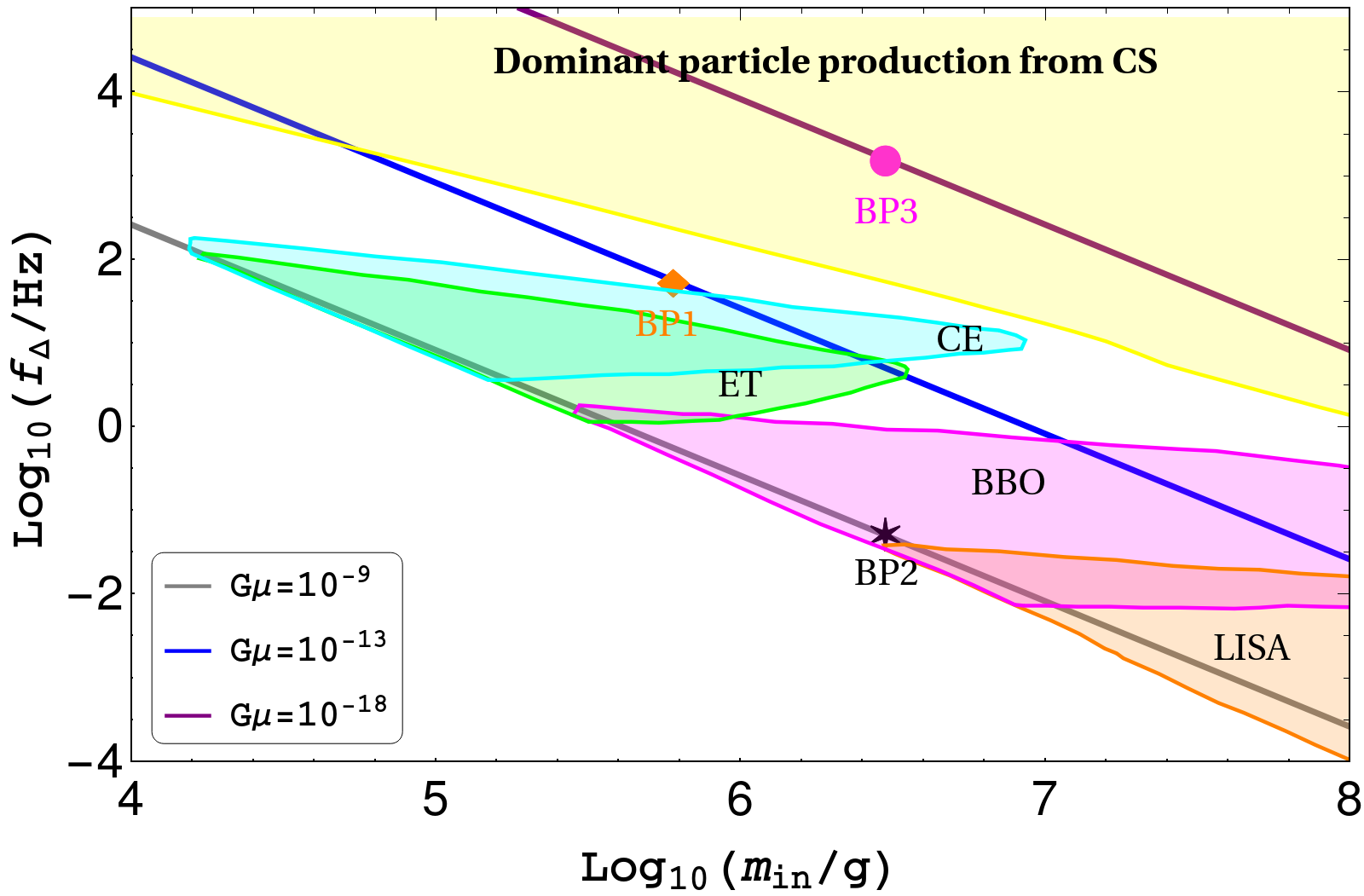}~~ \includegraphics[scale=.38]{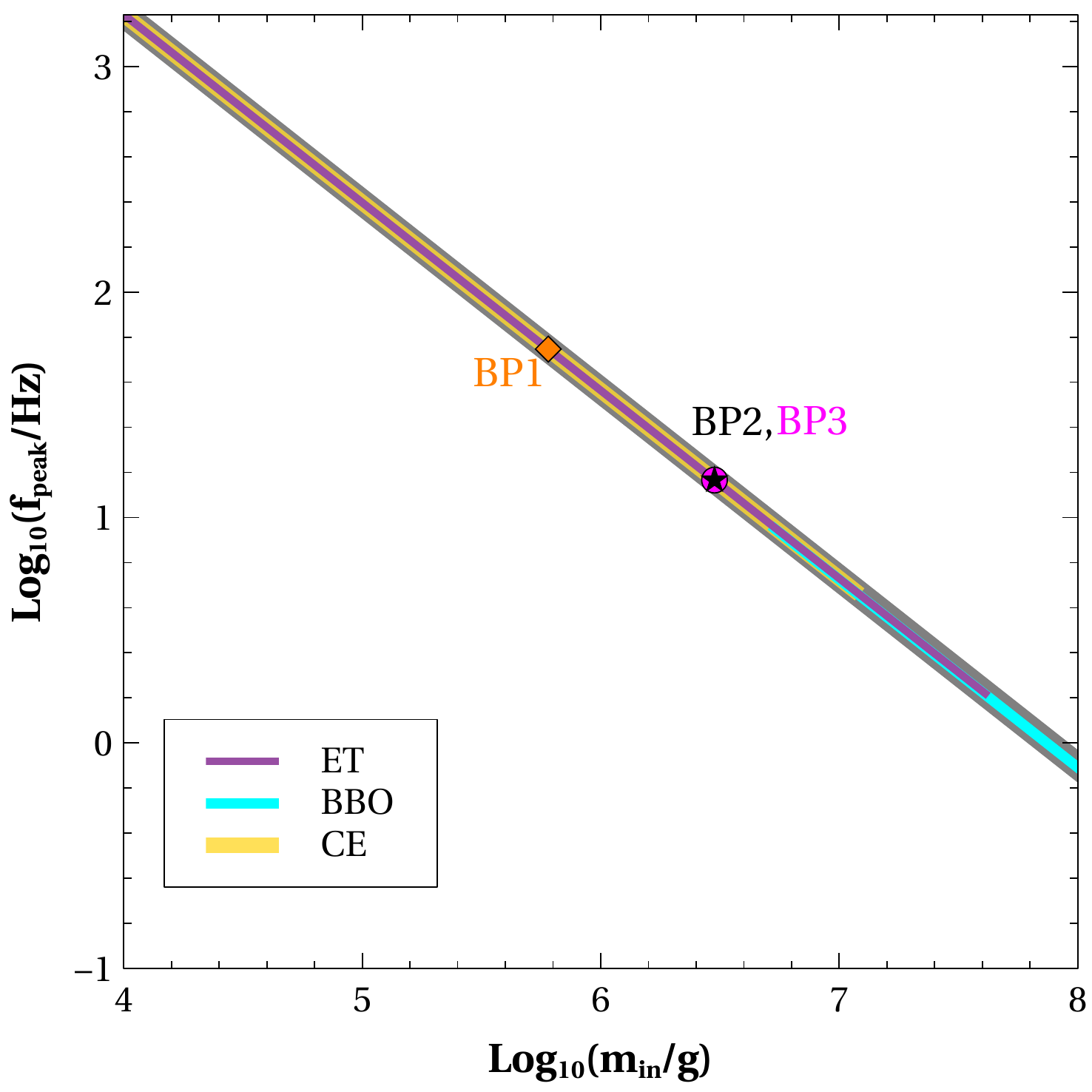}
    \caption{Left panel: Variation of $f_{\rm \Delta}$ with $m_{\rm in}$. In the yellow-shaded region, particle production from CS is dominant over GW radiation. Right panel: Variation of $f_{\rm peak}$ with $m_{\rm in}$. In both the plots, BP1, BP2, and BP3 (cf. Table \ref{BP}) are also shown with  $\blacklozenge$, $\bigstar$, and $\newmoon$ respectively. The sensitivities of different GW detectors are indicated too.}
    \label{fig4}
\end{figure}

 In Fig. \ref{fig1}, we show the combined spectrum corresponding to sub-cases (i) and (ii) for a fixed value of $G\mu = 10^{-13}$. The PBH parameters, $\beta, m_{\rm in}$, are chosen in such a way that the left panel plot corresponds to (i) $f_{\rm peak} > f_{\Delta}$ whereas the right panel shows the spectrum for (ii) $f_{\rm peak} < f_{\Delta}$. Each plot uses two different PBH benchmarks to show the peak position and amplitude shift. To show the variation due to the CS parameter $G\mu$, we plot the combined spectrum for different values of $G\mu$ in Fig. \ref{fig2}. The PBH parameters are kept fixed in both left and right panel plots of Fig. \ref{fig2},  whereas $G\mu$ values are chosen in such a way that left and right panels correspond to sub-case (i) and (ii), respectively. Finally, in Fig. \ref{fig3}, we show the sub-case (iii) $f_{\rm peak} \sim f_{\Delta}$ for different combinations of CS and PBH parameters\footnote{Note that another interesting frequency $f_{\rm eva}$ corresponding to the wave-number that crosses the horizon at $T_\Delta$ is, in general, different from $f_\Delta$ because $f_{\rm peak}\gg f_{\rm eva}$ \cite{Domenech:2020ssp}.}. In all these plots, the experimental sensitivities of SKA \cite{Weltman:2018zrl}, GAIA \cite{Garcia-Bellido:2021zgu}, EPTA \cite{Kramer_2013}, THEIA \cite{Garcia-Bellido:2021zgu}, $\mu$ARES \cite{Sesana:2019vho}, LISA\,\cite{2017arXiv170200786A}, DECIGO \cite{Kawamura:2006up}, BBO\,\cite{Yagi:2011wg}, ET\,\cite{Punturo_2010}, CE\,\cite{LIGOScientific:2016wof} and aLIGO \cite{LIGOScientific:2014pky} are shown as shaded regions of different colours. Clearly, depending upon the combination of $(G\mu, \beta, m_{\rm in})$ different parts of the spectrum--the plateau, peak, and the turning point frequency $f_{\Delta}$ remain within reach of different near future GW experiments. For convenience, we present a summary table including five benchmark points (those can also be found in the figures) and list the detectors with the potential to detect such unique GW signatures. Finally, in Fig.~\ref{fig4} (left panel), we show the variation of $f_\Delta$ with $m_{\rm in}$ for three different values of $G\mu$. In this plot,  the yellow-shaded triangular region in the upper right corner corresponds to the region where particle production from CS is more efficient than GW radiation \cite{Auclair:2019jip}. As mentioned earlier, this constraint comes from the critical size of the loops below which particle radiation becomes dominant. It  translates further to a lower bound on the initial time considered to compute the GW spectrum and introduces a break ($f_*$) in the spectrum similar to the early matter domination. Requiring that $f_*>f_\Delta$, for the loops containing cusps-like structures, the constraint on the parameter space can be derived as \cite{ Borah:2022byb} 
\bea
G\mu>2.4\times10^{-16}\left(\frac{T_\Delta}{\rm GeV}\right)^{4/5}.\label{partprod}
\eea
We translate this constraint to the $f_\Delta-m_{\rm in}$ plane using Eq.(\ref{eq:TPF}) and the evaporation temperature of PBH in the case where they dominate \cite{Datta:2020bht}.
In the right panel of Fig. ~\ref{fig4}, we show the variation of  $f_{\rm peak}$ with $m_{\rm in}$, which is independent of $G\mu$.  In both these plots, we indicate the sensitivities of different GW detectors and show three of the five benchmark points listed in Table ~\ref{BP}. As expected, in these plots, for BP1 ($G \mu=10^{-13}$) both $f_{\Delta}$ and $f_{\rm peak}$ fall within CE sensitivity whereas $f_{\rm peak}$ lies within ET sensitivity as well. On the other hand, BP2 ($G \mu=10^{-9}$) and BP3 ($G \mu=10^{-18}$) corresponds to the scenarios where $f_{\rm peak}>f_\Delta$ and $f_{\rm peak}<f_\Delta$ respectively. The analytical estimate of the turning point frequencies matches with our numerical analysis performed in generating Fig. \ref{fig1}- \ref{fig3}, up to a difference of less than $\mathcal{O} (1)$. 

\begin{table}[htb!]
\centering
\begin{tabular}{|c | c|c| c| c | c|}
\hline
BP   & $m_\text{in}$(g) & G$\mu$& $\beta$ &$f_{\rm peak}$ &$f_{\rm \Delta}$\\ \hline \hline
BP1 & $6 \times 10^{5}$ g & $10^{-13}$ & $2 \times 10^{-8}$ & ET, CE &     CE   \\ \hline
BP2 & $3 \times 10^{6}$ g  & $10^{-9}$ & $5 \times 10^{-9}$  & ET, CE &  BBO, DECIGO       \\ \hline
BP3 & $3 \times 10^{6}$ g & $10^{-18}$ & $5 \times 10^{-9}$  &ET, CE &   None      \\ \hline
BP4 & $4 \times 10^{5}$ g & $10^{-13}$ &$5 \times 10^{-8}$&LIGO, CE, ET  & CE\\ \hline
BP5 &$10^{8}$ g & $10^{-13}$  & $5 \times 10^{-10}$ & BBO, DECIGO &  LISA, BBO, DECIGO        \\ \hline
	\end{tabular}
	\caption{Benchmark points (BP) along with the GW detectors which can detect  the peak frequency $f_{\rm peak}$ and the turning point frequency $f_{\Delta}$ of the combined GW spectrum. }
	\label{BP}
\end{table}
{\it A brief summary of the detection prospect:} Consider that at low frequencies, we measure the stochastic GW background from cosmic strings (e.g., the recent finding of a stochastic common spectrum process by NANOGrav can be well explained by GWs from cosmic string for the parameter range $G\mu\in 10^{-11}-10^{-9}$ \cite{Goncharov:2021oub,Ellis:2020ena, Blasi:2020mfx,Samanta:2020cdk}). Then in Eq. (\ref{eq:TPF}), we have only one free parameter $T_\Delta$. In the case of PBH domination, this parameter is basically $m_{\rm in}$ since it appears as the only free parameter in $T_\Delta$. Therefore, if we observe a spectral break frequency $f_\Delta$ at any high-frequency detector, e.g., ET and CE, we expect another sharp peak at a ``predicted frequency ($f_{\rm peak}$)" according to Eq. (\ref{eqn:fpkpbh}), or vice versa. Such a complementary detection prospect distinguishes the PBH-dominated epoch from any other EMDs. In this context, let us highlight the consequence of summing over all the modes while computing the GW spectrum from cosmic strings, especially for $f_{\rm peak}>f_\Delta$. Because summing over large $k$-modes changes the slope of the spectrum  from $f^{-1}$ to $f^{-1/3}$ (loops containing cusps), such a $f_\Delta-f_{\rm peak}$ complementary detection prospect requires a longer period of PBH domination (large $\beta$, cf. Eq. \eqref{eqn:omgpeakpbh}) compared to the simplest $k=1$ case, so that $\Omega_{\rm gw}^{\rm peak}(f_{\rm peak}>f_\Delta)\gg\Omega_{\rm gw}^{\rm CS}(f_{\rm peak}>f_\Delta)$.\\

{\it Connection to realistic BSM scenarios:} Given the general discussion above, let us point out how such a spectrum could be a probe of particle physics models. First, although the PBHs eventually evaporate, they might produce stable or unstable relics. A stable relic could be dark matter, whereas an unstable particle (such as right-handed neutrinos in seesaw mechanism) produced from PBH may seed  baryogenesis (see, e.g., Refs. \cite{pbhbary,pbhdm,rel1,rel2}). In addition, depending on the mass and relative energy fraction, a PBH-dominated universe can alter the standard parameter space of many high-energy models, e.g., dark matter models \cite{Gondolo:2020uqv}, and the models of light QCD-axions \cite{Bernal:2021yyb}. Thus PBHs (the mass and the energy fraction) act as a portal between gravitational waves and the parameters of high-energy particle physics models, motivating a synergic search of those models with GWs plus the conventional particle physics experiments. Such models featuring a high-scale gauged $U(1)$-symmetry breaking would exhibit the combined spectrum discussed in this article. See, e.g., recent work on the seesaw mechanism with a gauged $U(1)$-symmetry \cite{Borah:2022iym}, where the motivation and consequences of PBH domination have been studied. Studying this framework in the context of global strings, which generically appear in QCD-axion (including axion-like particles) models, might be interesting to explore in future works. \\

Before concluding, let us stress that the motivation of this paper is to point out that the consequences of an EMD due to the standard long-lived fields or some additional new physics and PBHs could be vastly different. In particular, distorting GWs spectrum from cosmic strings with EMD, results in a scale-invariant spectrum followed by a red tilt in both cases, while for the latter one,  depending on the initial energy fraction of PBHs, a sharp blue tilt might follow or precede. This makes the PBH scenarios unique. More interestingly, depending on the model parameters, the new features ($f_\Delta$ and $f_{\rm peak}$) in the scale-invariant spectrum might appear in a single detector (cf. Fig.\ref{fig3}). Additionally, because the effects of ultra-light PBHs are of great interest to study, e.g., in the context of dark matter production, baryogenesis, and axions, such spectral features in the GWs could be a unique probe of these models, irrespective of the particle physics coupling strength. Let us also mention another interesting aspect of the PBH-string scenario. We focus on the frequency range where dominant GWs come from the post-PBH evaporation era, i.e., from the beginning of the most recent  radiation domination. Although we do not discuss the detailed spectral shapes of the  high-frequency GWs that might arise from the loops produced during PBH domination  and before, an additional effect might be at play, distorting the high-frequency part of the spectrum.  This effect is black hole-string interaction which has been studied analytically in the context of long-lived black holes, contributing to the dark matter density \cite{Vilenkin:2018zol}. The average separation of PBHs after formation  is $d\simeq \lambda^{-1/3} t_f$, where $\lambda\ll 1$ is the fraction of horizon that turns into PBHs. In this case, long strings of average length $l\simeq t_f/\lambda$ and interaction probability $\lambda$ \cite{Vilenkin:2018zol} connecting two BHs  can randomly chop off loops less than the size of the horizon ($\alpha\simeq 0.1$) $t\sim t_f$. Therefore, the  string dynamics are largely unaffected by the PBH during the initial stages of evolution after PBH formation, and the usual computation of GWs from cosmic string loops holds. Nonetheless, the efficiency of chopping off loops of horizon size or less reduces as time evolves when the PBH separation connecting long strings becomes comparable to the horizon, and two PBH tend to straighten long strings. The exact quantification of such a phenomenon requires appropriate numerical simulation.

\section{Conclusion}
\label{sec4}
We have proposed a unique gravitational wave-based probe of super-high scale $U(1)$ symmetry breaking with a PBH-dominated epoch before the BBN era. While cosmic strings resulting from $U(1)$ breaking lead to a typical scale-invariant GW spectrum, ultra-light PBH domination leads to an additional observable GW spectrum from density fluctuations. When combined, the GW spectrum has a unique shape with a plateau, a sharp tilted peak, and a characteristic fall-off behaviour. Depending on the cosmic string and the PBH parameters, different parts of the spectrum fall within reach of ongoing and planned future experiments. While early matter domination, e.g., by a scalar field or supercooled first-order phase transition along super-high scale $U(1)$ symmetry breaking can lead to  a flat plateau followed by a fall-off, peak over a flat plateau or combination of both, it remains distinguishable from the spectrum we obtain in the presence of PBH. This is due to the lower and upper bounds on peak frequency in our work due to the allowed ultra-light PBH mass window, which does not exist in other scenarios,  plus the distinct power-law behaviour of the GWs from PBH density fluctuations. In addition to such marked features verifiable in GW detectors, the setup discussed in our work can also have very rich phenomenological implications connected to the production of dark matter from PBH evaporation, high-scale leptogenesis, and seesaw for neutrino mass related to $U(1)$ symmetry breaking \cite{Barman:2022gjo, Borah:2022vsu, Datta:2020bht, Samanta:2021mdm, Borah:2022iym}.

\section*{Acknowledgements}
The work of D.B. is supported by SERB, Government of India grant MTR/2022/000575. R.S. acknowledges the European
Structural and Investment Funds and the Czech Ministry of Education, Youth and Sports
(Project CoGraDS -CZ.02.1.01/0.0/0.0/15003/0000437) for the financial support. R.R. acknowledges the National Research Foundation of Korea (NRF) grant
funded by the Korean government (2022R1A5A1030700) and also the support provided by the Department of Physics, Kyungpook National University.

\providecommand{\href}[2]{#2}\begingroup\raggedright\endgroup

\end{document}